\documentstyle[psfig]{lamuphys}
\makeatletter
\let\chapter\hid@chapter
\makeatother
\newcommand{\kms}{{km s$^{-1}$}\/}
\newcommand{\as}{{$''\!\!$.}\/}
\newcommand{\HST}{{\sl HST}\/}
\newcommand{\HaNii}{{H$\alpha$+[{\sc N ii}]}\/}
\begin{document}

\index{spectrum!nuclear}
\index{variability!nuclear}

\pagenumbering{arabic}
\title{Morphology of the Nuclear Disk in M87}

\author{Z.I.\,Tsvetanov\inst{1}, 
        M.G.\,Allen\inst{1,2},
        H.C.\,Ford\inst{1,3}, and
        R.J.\,Harms\inst{4}}

\institute{Johns Hopkins University, Baltimore, MD 21218, USA
\and
Mount Stromlo and Siding Spring Observatories, ACT 2611, Australia
\and
Space Telescope Science Institute, Baltimore, MD 21218, USA
\and
RJH Scientific, 5904 Richmond Highway, Alexandria, VA 22303, USA}

\maketitle

\begin{abstract}

A deep, fuly sampled diffraction limited (FWHM $\sim$ 70 mas)
narrow-band image of the central region in M87 was obtained with the
Wide Filed and Planetary Camera 2 of the {\it Hubble Space Telescope}
using the dithering technique.  The \HaNii\ continuum subtracted image
reveals a wealth of details in the gaseous disk structure described
earlier by Ford et al.\ (1994). The disk morphology is dominated by a
well defined three-arm spiral pattern. In addition, the major spiral
arms contain a large number of small ``arclets'' covering a range of
sizes (0\as1--0\as3 = 10--30 pc). The overall surface brightness
profile inside a radius $\sim$ 1\farcs5 (100 pc) is well represented
by a power-law $I(\mu) \sim \mu^{-1.75}$, but when the central $\sim$
40 pc are excluded it can be equally well fit by an exponential
disk. The major axis position angle remains constant at about PA$_{\rm
disk} \sim 6^{\circ}$ for the innermost $\sim 1''$, implying the disk
is oriented nearly perpendicular to the synchrotron jet (PA$_{\rm jet}
\sim 291^{\circ}$). At larger radial distances the isophotes twist,
reflecting the gas distribution in the filaments connecting to the
disk outskirts. The ellipticity within the same radial range is $e =
0.2-0.4$, which implies an inclination angle of $i \sim 35^{\circ}$.
The sense of rotation combined with the dust obscuration pattern
indicate that the spiral arms are trailing.

\end{abstract}

\section{Introduction}

The disk of ionized gas in the nucleus of M87 is currently the best
example of a family of similar small ($r \sim 100$ pc) gaseous disks
found to be common in the centers of elliptical galaxies with active
nuclei (for a review see Ford et al.\ 1998). Several \HST\ kinematical
studies have shown that in M87 the gas is in Keplerian rotation,
orbiting a massive black hole with a mass $M_{\rm BH} \sim 2 - 3
\times 10^9 M_{\odot}$ (Harms et al.\ 1994; hereafter H94, Ford et
al.\ 1996a,b; and Macchetto et al.\ 1997, hereafter M97). The few
other galaxies studied kinematically so far (NGC 4261 -- Ferrarese et
al.\ 1996, NGC 6521 -- Ferrrarese et al.\ 1998, NGC 4374 -- Bower et
al.\ 1998) have further shown that nuclear gaseous disks offer an
excellent tool for measuring the central black hole mass.

Recent studies have revealed other important characteristics of the
nuclear disk in M87. Its aparent minor axis (F96, M97) is closely
aligned with the synchrotron jet ($\Delta\theta \sim 10^{\circ} -
15^{\circ}$) suggesting a causal relationship between the disk and the
jet. The system of filaments in the center of M87 (Sparks, Ford \&
Kinney 1993; SFK) may also be causally connected to the disk. For
example, the filaments extending $\sim$17$''$ (1200 pc) to the NW at
PA $\sim 315^{\circ}$ are blue shifted with respect to systemic
velocity and show dust absorption implying they are on the near side
of M87 as is the jet. These two findings led SFK to conclude that
these filaments are streamers of gas flowing away from the center of
M87 rather then falling into it. The images in F94 (see also Ford \&
Tsvetanov, this volume, FT98) show an apparent connection between at
least some of the larger scale fillaments and the ionized nuclear
disk.

Direct spectroscopic evidence for an outflow was found recently.
Several UV and optical absorption lines from neutral and very mildly
ionized gas were measured in the FOS spectrum of the nucleus
(Tsvetanov et al.\ 1998; T98). These lines are broad (FWHM $\sim 400$
\kms) and blue shifted by $\sim 150$ \kms\ with respect to M87's
systemic velocity implying both an outflow and turbulence. In
addition, non-circular velocity components -- both blue and red
shifted -- were found at several locations in the disk (F96, FT98),
and observed emission lines are much broader than the expected
broadening due to the Keplerian motion accross the FOS aperture. All
these properties are best understood if a bi-directional wind from the
disk were present. This wind may be an important mechanism for
removing angular momentum from the disk to allow accretion through the
disk onto the central black hole.

Whatever the physical conditions in the disk it is important to map
its morphology in detail. The first \HST\ images (F94) have hinted
that a spiral pattern could be present, but the signal-to-noise was
too low for a definitive conclusion.  In this paper we present deep,
fully sampled diffraction limited narrow band images of the nuclear
region in M87.  We use these images to characterize the ellipticity,
brightness distribution, and morphology of the disk. In this paper we
adopt a distance to M87 of 15 Mpc, corresponding to a scale of 1$''$ =
73 pc.

\vspace{-2mm}
\section{Observations}
\vspace{-2mm}

M87 was imaged though the WFPC2 narrow-band filter F658N (central
wavelength/effective width = 6590/28 \AA) which isolates the \HaNii\
emission line complex at the systemic redshift. The nucleus was placed
near the center of the planetary camera (PC) and the \HST\ roll angle
was 307$^{\circ}$, orienting the M87 synchrotron jet roughly along the
diagonal of the PC field of view.
 
\begin{figure}[ht]
\vspace{-15mm}
\centerline{\psfig{figure=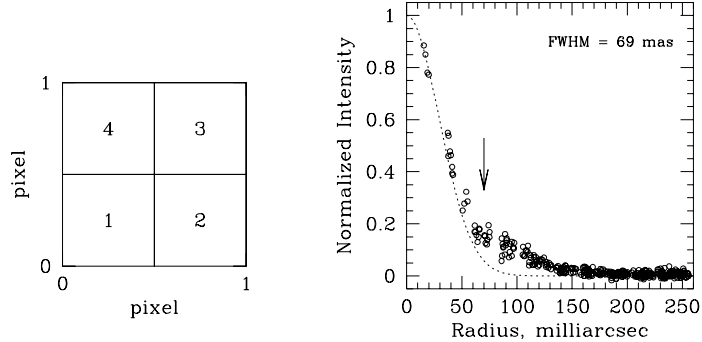,width=10.0cm}}
\vspace{-35mm}
\caption{{\it Left} panel illustrates the dithering pattern.  Three
images were taken at each of the positions numbered 1 to 4. The
telescope offset was 5.5 PC1 pixels between positions. The {\it right}
panel shows the radial profile of an unsaturated star $\sim$6$''$
south of nucleus.  The overplotted Gaussian has a FWHM equal to the
\HST\ diffraction limit at the wavelength of observation (6580
\AA). The first Airy ring is marked by the
arrow. \label{fig:dith-patt+star-prof} }
\vspace{-3mm}
\end{figure}

The PC has a scale of 45.54 mas pixel$^{-1}$ (Holtzman et al.\ 1995)
and undersamples the \HST\ diffraction limited PSF at all wavelengths
shorter of $\sim 1\mu$. To achieve full sampling we took images at
four adjacent positions, offsetting the telescope by 5.5 PC pixels
between each images -- the so called dithering technique.  The pattern
used is illustrated in Fig.~\ref{fig:dith-patt+star-prof}.  At each of
the four subpixel positions we took 3 images for cosmic ray rejection
and cleaning of permanent defects.  We re-calibrated the images using
the best recommended calibration files.
 

To build a fully sampled image we used the variable-pixel linear
reconstruction algorithm (informally known as ``drizzling'') developed
originally for the Hubble Deep Field (Fruchter \& Hook 1997).  The
relative offsets of individual images were estimated from the
positions of good signal-to-noise objects in the field and also
through a 2-D cross correlation. Both techniques yielded similar
results.  No significant offsets were found for the three frames taken
at each one of the subpixel positions. These were combined to remove
the cosmic ray events. The four combined frames were then combined on
an output grid of 25 mas pixel$^{-1}$ using the drizzling
algorithm. The 25 mas pixel size is a good match for the $\sim$ 70 mas
\HST\ diffraction limit at H$\alpha$. This is illustrated by the
radial profile of a star in the field shown in
Fig.~\ref{fig:dith-patt+star-prof} where the first Airy ring is
clearly seen.
 
Because of serious observing time restrictions, and also because the
underlying galaxy profile is known to be smooth, no continuum images
were obtained simultaneously with the on-line ones. To create a
suitable off-band image for continuum subtraction we used a number of
M87 images from the \HST\ public archive. These include a series of
F814W images taken on November 11, 1995, which have small realtive
offsets that allow us to build a fully sampled image using the same
drizzling technique. The offset pattern in not ideal (as in the case
of F658N), but this is less critical because of the smooth shape of
the continuum light distribution.

To form an image suitable for continuum subtraction at the wavelength
of H$\alpha$ we used a ($V-I$) color image to correct the F814W
image for color effects, which are particularly important at the
positions of the dust bands. Finally, the continuum subtracted \HaNii\
image was flux calibrated using the estimated system throughput at the 
redshifted position of H$\alpha$ and the averaged mesured 
[{\sc N ii}]/H$\alpha$ line ratio.


\vspace{-2mm}
\section{Disk morphology}
\vspace{-1mm}
\index{M87!nuclear disk!morphology}

\begin{figure}[ht]
\vspace{-44mm}
\centerline{\psfig{figure=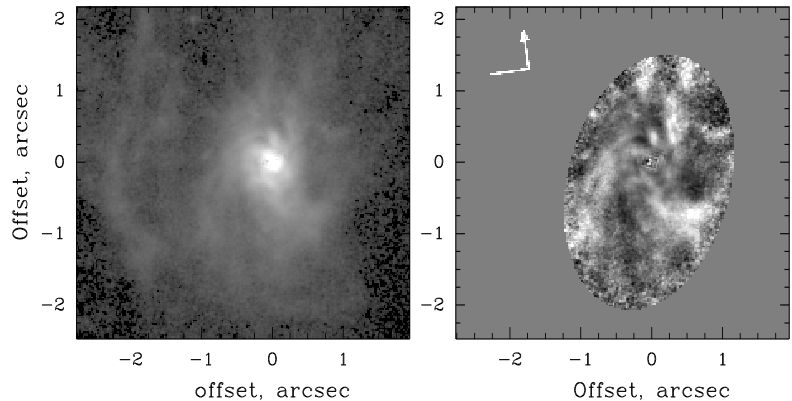,width=12.0cm}}
\vspace{-60mm}
\caption{The \HaNii\ continuum subtracted image is shown in the {\it
left} panel. The gray scale saturates the central few pixels, which
are uncertain due to the imperfect color matching and variability of
the central point source. The {\it right} panel shows the ratio of the
observed structure to the smooth disk model, as duscussed in the
text. The image is stretched linearly between $\pm$70\% relative to
the model. \label{fig:disk+spiral-patt} }
\vspace{-3mm}
\end{figure}

The central $1''-2''$ of the continuum subtracted \HaNii\ image
(Fig.~\ref{fig:disk+spiral-patt}, left panel) is dominated by a
clockwise winding 3-arm spiral pattern superposed on an underlying
disk-like morphology. At larger radial distances the gaseous structure
is less well organized, asymmetric, and gradually transition into the
larger scale H$\alpha$ filaments.  Several of these filaments can be
traced to connect to the disk but none appears to go directly into the
nucleus.

To estimate the parameters describing the surface brightness profile,
ellipticity and orientation, we have performed an elliptical isophote
analysis. We started by setting all parameters free, and gradually
introduced constraints such as fixed center position, major axis
position angle and/or ellipticity, in an attempt to explore the
robustness of the solution. To our satisfaction the values for all
major parameters of the fit remained stable regardless of the
constraints used.

\begin{figure}[ht]
\vspace{-2mm}
\centerline{\psfig{figure=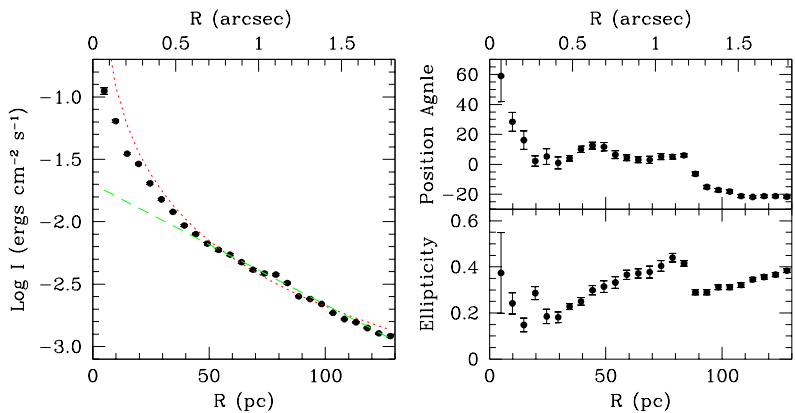,width=11.0cm}}
\vspace{-52mm}
\caption{{\it Left:} The surface brightness profile of the M87 nuclear
gaseous disk. The short dashed line is a power law $I_{\rm pl} \sim
(r/50)^{-1.75}$, and the long dahsed line is an exponential disk with
$I_{\rm disk} \sim exp(-r/45)$, where $r$ is in pc.  {\it Right:} The
radial dependence of the major axis position angle (PA$_{\rm maj}$)
and ellipticity ($e$). \label{fig:PAmaj+e} }
\vspace{-3mm}
\end{figure}

Fig.~\ref{fig:PAmaj+e} shows the surface brightness radial profile,
major axis position angle, and ellipticity as a function of radial
distance. Outside $r \sim$ 0\as5 (40 pc), the surface brightness
profile is well described by an exponential disk with an effective
radius of 45 pc. There is, however, a significant excess of light in
the central 0\as5, and a single power law $I_{\rm pl} \sim
(r/50)^{-1.75}$ is a better representation of the overall profile.  In
the region 20--90 pc, the orientation of the major axis remains
roughly constant at PA$_{\rm maj} \sim 6^{\circ}$. At the same time
the ellipticity increases smoothly from 0.2 to 0.4. The constancy of
PA$_{\rm maj}$ is usually interpreted as a signature of a disk, while
the ellipticity could be influenced by the spiral arms or there could
be a warp. The central few points are less reliable because of
possible color mismatch of the on- and off-band images and the
variability of the nuclear point source (T98).

The kinematic studies (F96, M97) have shown that the north part of the
disk is receiding and the south part is approaching and the estimated
inclination is consistent with the one inferred from isophote
analysis.  The obscuration patches on the SE side (F94) indicate that
the SE side is closer to us. Combining all this information indicates
that the spiral arms are trailing.

This work is supported by NASA grant NAG-1640 to the HST FOS team.

%
%

\end{document}